# Peculiarities of geomagnetic field behavior near boundary of the mesozoic - cenozoic

A. Yu. Kurazhkovskii[1], N. A. Kurazhkovskaya[2], B.I. Klain[3]

[1, 2, 3]*Geophysical Observatory Borok, Schmidt Institute of Physics of the Earth of the Russian Academy of Sciences, Borok, Yaroslavl Region, 152742, Russian Federation*
[1]E-mail:*ksasha@borok.yar.ru*
[2]E-mail:*knady@borok.yar.ru*
[3]E-mail:*klain@borok.yar.ru*

**Annotation.** The results of paleointensity determinations of the geomagnetic field, which are obtained by sedimentary rocks in the Middle Jurassic - Neogene (167-12) Ma, are summarized. It was found that in the interval (167–67) Ma (Middle Jurassic - Cretaceous) mean values and amplitudes of paleointensity variations increased. Mean values of paleointensity reached a maximum in front of the Cretaceous-Paleogene boundary. During the Paleogene the mean values and amplitudes of paleointensity variations decreased. In the Cretaceous - the beginning of the Paleogene cyclic changes in the amplitudes of paleointensity variations with characteristic times of several million years have been observed. It is shown that the duration of cycles of the paleointensity changes coincided with transgressive-regressive cycles (T-R cycles). This can be considered as a confirmation of the hypothesis on a relation between processes in the Earth's core and tectonic processes.

1. **Introduction**

The behavior of the characteristics of the ancient geomagnetic field (the frequency of geomagnetic inversions and paleointensity) is associated with the dynamics of convection processes in the core of the Earth. In accordance with this postulate paleomagnetic data are used to study of the relation between processes in the core of the Earth and processes on its surface. So, changes in the characteristics of the geomagnetic field during geotectonic cycles (Bertrand cycles by duration of about 200 Ma) were found, for example, [*Khramov et al.*, 1982; *Petrova et al.*, 2000]. A study of the Pleistocene paleointensity behavior showed that its variations have much less characteristic times than the cycles of Bertrand [*Guyodo and Valet*, 1999]. However, studies of the paleointensity variations at large intervals of geological time, for example, the Mesozoic and Cenozoic have not yet been carried out. To study the changes in the intensity of thermal processes in the core of the Earth detailed paleointensity data, which can be obtained from ancient sedimentary rocks, is needed. Similar studies of the paleointensity began only in recent years [*Cronin et al.*, 2001; *Ohneiser et al.*, 2013; *Yamamoto et al.*, 2014; *Kurazhkovskii et al.*, 2015a; 2015b; 2016].

In the present study we summarized these data and the behavior of paleointensity on a large interval of geological time (167 – 12) Ma were obtained. On this basis, the correspondence of paleointensity behavior to the most known global geological events – to the change of geological eras and transgressive-regressive cycles (T-R cycles) has been investigated.

**2. Analyzed data**

The curve characterizing the behavior of the paleointensity is based on the results of its definitions presented in [*Kurazhkovskii et al.*, 2015a; 2015b; 2016] (167 to 23 Ma), [*Cronin et al.*, 2001] (99 to 85 Ma), [*Yamamoto et al.*, 2014] (40 to 23 Ma), and [*Ohneiser et al.*, 2013] (19 to 12 Ma). To synthesize these paleointensity data, we calibrated them. The calibration of the author's data was carried out according to the method in which the redeposition was used [*Kurazhkovskii et al.*, 2011]. Calibration of the results of the paleointensity determinations from the works [*Cronin et al.*, 2001; *Yamamoto et al.*, 2014; *Ohneiser et al.*, 2013] was carried out using the results of paleointensity determinations by thermomagnetized rocks from the database PINT(2015.05) [http://earth.liv.ac.uk/pint/]. The description of this database is given by Biggin et al., (2010).

We used data on eustatic ocean level changes or T-R cycles from [*Haq et al.*, 1988] and from the Geologic Time Scale 2008 (GTS 2008) [Gradstein et al., 2008] for comparison of the behavior of the paleointensity and the cyclicity of global tectonic processes. T-R cycles are a reflection of the global cyclicity of rifting-folding [*Milanovsky*, 1996].

**3. Behavior of the paleointensity**

The summarized data of the paleointensity determinations obtained by sedimentary rocks are shown in Fig. 1a. A smoothed curve of the paleointensity behavior is presented in the Fig. 1b. As can be seen from Fig. 1, the amplitude of the paleointensity variations and its mean values increased from the Middle Jurassic to the end of the Cretaceous. The most significant increase in the amplitude of the paleointensity variations occurred at the end of the Jurassic - the beginning of the Cretaceous and at the end of the Cretaceous. Then, during the Paleogene the mean values of the paleointensity and the amplitude of its variations decreased. Thus, at the boundary of the geological eras (Mesozoic - Cenozoic) turning-point was occurred in the behavior of the paleointensity (the growth of paleointensity was replaced by it's a decrease).

As can be seen from Fig. 1a, in the Jurassic-Paleogene cyclic changes in the amplitudes of the paleointensity variations (alternation of paleointensity variations of large and small amplitude) took place. The characteristic times of such changes in paleointensity were several million years (an average of 6 million years). Most clearly such cyclicity in the paleointensity behavior was also revealed near the Mesozoic - Cenozoic boundary.

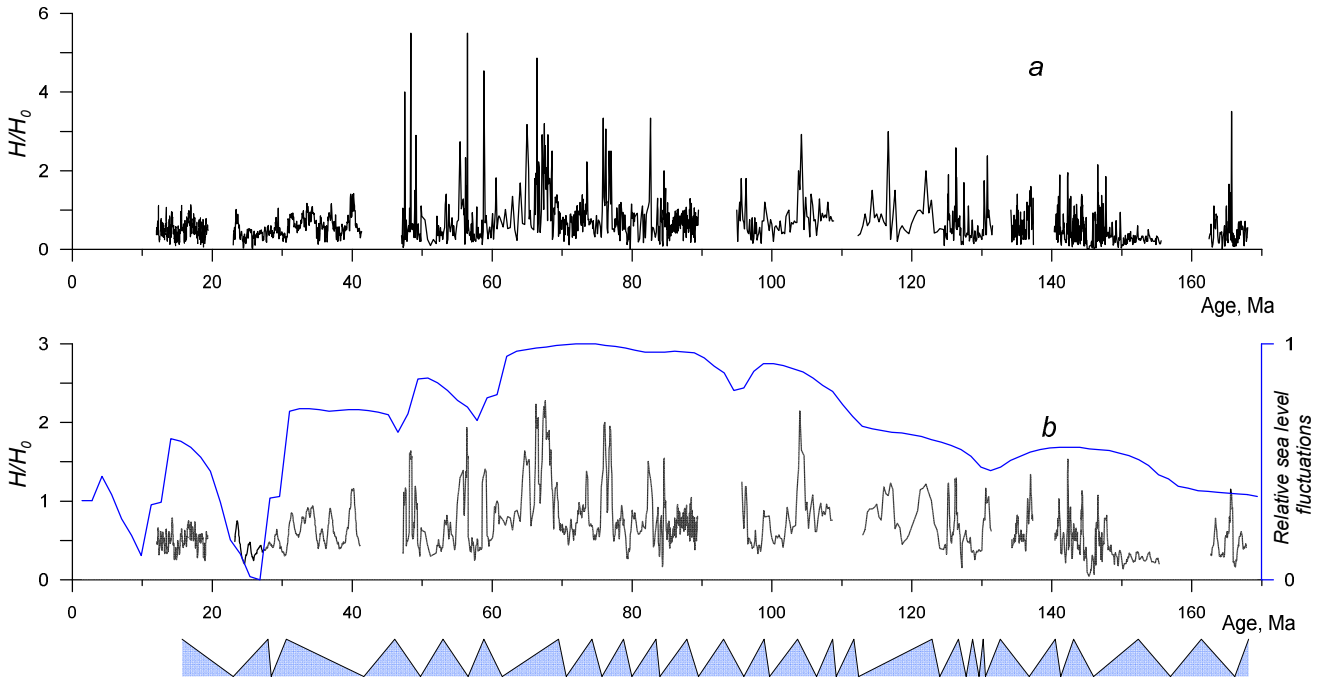

Fig. 1. (a) Behavior of the paleointensity in the interval (167-12) Ma, (b) smoothed paleointensity data by sedimentary rocks. The blue line shows relative changes in sea level according to Haq et al., (1988). The data of T-R cycles of second-orders (blue triangles) from [*Gradstein et al.*, 2008] are shown below the abscissa axis.

The results of the paleointensity determinations by sediments allow us to compare its behavior with the T-R cycles. As it seen from Fig. 1b, changes in the paleointensity and ocean level [*Haq et al.*, 1988] coincide. According to GTS 2008, the average duration of T-R cycles of second-order (as well as variations of the paleointensity) was 6 million years. The uncertainty of the stratigraphic dating, on the basis of which associating of the paleomagnetic data to the GTS 2008 scale was carried out, can be several hundred thousand years (up to 1 million years) [*Guzhikov and Baraboshkin*, 2006]. This does not make it possible to determine the exact phase correspondence of the paleointensity variations to T-R cycles. Nevertheless, identical characteristic times of changes of the paleointensity and T-R cycles indicate a connection between the processes in the liquid core and the lithosphere.

**3. Discussion**

The paleointensity data obtained by sedimentary rocks allowed detecting the series of earlier unknown regularities in the behavior of the ancient geomagnetic field. So, the change of geological eras is usually associated with accidental catastrophic events on the Earth's surface (the fall of a large meteorite, the activation of magmatism). Analysis of the paleointensity data showed that on the boundary of the Mesozoic and Cenozoic, the processes in the Earth's core were reformed. This is evidenced by the behavior of paleointensity (Fig. 1b). Thus, with the change of geological eras, changes both in the outer (biosphere and lithosphere) and in the inner (core, mantle) shells of the Earth occurred.

The coincidence of the characteristic times of paleointensity variations and T-R cycles can be an indirect confirmation of the hypothesis of temperature pulsations and (as a consequence) the volume of the liquid core of the Earth. Earlier arguments in favor of the validity of this hypothesis were considered by *Milanovsky*, (1996). In this study the relationship between changes in the frequency of geomagnetic inversions, T-R cycles and rifting-folding cycles was analyzed. According to *Milanovsky*, (1996) changes in the frequency of geomagnetic field inversions are associated with variations of the temperature of the liquid core. It was assumed that the growth of the core temperature coincides with the phases of the Earth's expansion, activations of rifting and transgressions. The relationship between the intensity of the geomagnetic field and the intensity of thermal processes in the liquid core follows from the ideas about the work of geodynamo, for example, [*Jones,* 2011; *Reshetnyak and Pavlov*, 2016]. Geodynamo models and the materials of this study provide a basis for the assumption that there is a connection between the changes in the temperature of the liquid core with variations in the paleointensity and processes in the lithosphere. Pulsations of temperature and volume of the Earth's core can occur due to the presence of heat sources and low thermal conductivity of the mantle [*Romashov*, 2003]. An increase of the core temperature is accompanied by an adiabatic expansion, which under the influence of gravity is replaced by compression. Thus, the Earth is considered as an oscillatory system [*Romashov*, 2003]. In such a system, the connection of processes in the core of the Earth with processes on its surface can be carried out almost instantaneously.

## 4. Conclusions

In the interval (167-67) Ma (Middle Jurassic- Cretaceous) mean values and amplitudes of the paleointensity variations were increasing. Mean values of paleointensity reached a maximum in front of the Cretaceous-Paleogene boundary. Then, during the Paleogene the mean values and amplitudes of the paleointensity variations decreased. Thus, the Mesozoic-Cenozoic boundary is marked by a turning-point in behavior of the paleointensity. In the Cretaceous - the beginning of the Paleogene cyclic changes in the amplitudes of paleointensity variations with characteristic times of several million years have been observed. The average durations of cycles of paleointensity changes and of transgression-regression cycles coincided.

**References**

Biggin, A.J., McCormack, A., Roberts, A., 2010. Paleointensity database updated and upgraded. *EOS Trans. Am. geophys. Un.*, 91, 15, doi:10.1029/2010EO020003.

Cronin, M., Tauxe, L., Constable, C., Selkin, P., Pick, T. 2001. Noise in the quiet zone. *Earth and Planet. Sci. Lett.*, 190, 13–30.

Gradstein, F.M., Ogg, G.J., van Kranendonk, M**.,** 2008. On the Geologic Time Scale 2008. *News letters on stratigraphy*, 43/1, 5–13.

Guzhikov, A.Yu., Baraboshkin, E.J., 2006. Assessment of diachronism of biostratigraphic boundaries by magnetochronological calibration of zonal scales for the lower cretaceous of the Tethyan and Boreal belts. *Doklady Earth Sciences*, 409, 6, 843–846.

Guyodo, Y., and Valet, Y.-P., 1999. Global changes in intensity of the Earth's magnetic field during the past 800 kyr. *Nature*, 399, 249–252.

Haq, B.U., Hardenbol, J., Vail, P.R., 1988. *Mesozoic and cenozoic chronostratigraphy and cycles of sea-level changes.* Sea-level changes: an integrated approach. Soc. Econ. Paleontologists and Mineralogists. Spec. Puble. USA, Oclahoma: Tulsa, 42, pp. 11-108.

Jones, C., 2011. Planetary magnetic fields and fluid dynamos. *Annual Review of Fluid Mechanics*, 43, 583–614.

Khramov, A.N., Goncharov, G.I., Komissarova, R.A., Pisarevskii, S.A., Pogarskaya, I.A., Rzhevskii, Yu.S., Rodionov, V.P., and Slautsitais, I.A., 1982. *Paleomagnetology* (in Russian), Khramov, A.N., Ed., Leningrad: Nedra, pp. 312.

Kurazhkovskii, A.Yu., Kurazhkovskaya, N.A., Klain, B.I., 2011. Calibration of geomagnetic paleointensity data based on redeposition of sedimentary rocks. *Physics of the Earth and Planetary Interiors*, 189, 1-2, 109–116, doi: 10.1016/j.pepi.2011.08.002.

Kurazhkovskii, A.Yu., Kurazhkovskaya, N.A., Klain, B.I., 2015a. Stochastic behavior of geomagnetic field in the Middle Jurassic – Paleogene. *Geomagnetism and Aeronomy*, 55, 2, 223–234. doi: 10.1134/S0016793215010089.

Kurazhkovskii, A.Yu., Kurazhkovskaya, N.A., Bagaeva, M.I., Guzhikova, A.A., 2015b. Possible Changes of the Geomagnetic Field Intention in Titonian – Berrassian and Campanian – Maastrichtian. *Izvestiya of Saratov University. New series. Serie*s: *Earth Sciences*, 15, 2, 41–46.

Kurazhkovskii, A.Yu., Kurazhkovskaya, N.A., Surinskii, A.M., 2016. Definition of Paleointensity in the Eocene Section Plateau Aktolagay. *Izvestiya of Saratov University. New series. Serie*s: *Earth Sciences*, 16, 3, 172–178.

Milanovsky, E. E., 1996. The Correlation between Higher Frequency Phases of Geomagnetic Reversals, Drops in Sea Level, and Crustal Compressive Deformations in the Mesozoic and Cenozoic. *Geotectonics*, 30, 1, 1–8.

Ohneiser, C., Acton, G., Channell, J.E.T., Wilson, G.S., Yamamoto, Y., Yamazaki, T., 2013. A middle Miocene relative paleointensity record from the Equatorial Pacific. *Earth and Planet. Sci. Lett.*, 374, 227–238.

Petrova, G.N., Pechersky, D.M., Khramov, A.N., 2000. Paleomagnetology is the science of the 20th century. *Izvestiya. Physics of the Solid Earth*, 36, 9, 777–798.

Reshetnyak, M.Y., Pavlov, V.E., 2016. Evolution of the dipole geomagnetic field. Observations and models. *Geomagnetism and Aeronomy*, 56, 1, 110–124. doi: 10.1134/S0016793215060122.
doi: 10.1134/S0016793215060122.

Romashov, A.N., 2003. *Planet Earth: tectonophysics and evolution* (in Russian). Moscow: Editorial URSS, pp. 364.

Yamamoto, Y., Yamazaki, T., Acton, G.D., Richter, C., Guidry, E.P., Ohneiser C., 2014. Palaeomagnetic study of IODP Sites U1331 and U1332 in the equatorial Pacific - extending relative geomagnetic palaeointensity observations through the Oligocene and into the Eocene. *Geophys. J. Int.*, 196(2), 694–711.